\documentclass[final,5p,times,twocolumn]{elsarticle}

\usepackage{lineno}
\usepackage[hidelinks]{hyperref}

\usepackage{graphicx}
\usepackage{amsmath}  
\usepackage{xspace}

\def\mum{\ensuremath{\mu}m}

\begin{document}

\begin{frontmatter}

\title{Large area curved silicon modules for future trackers}

\author[2,3]{S. Moss}

\author[2]{Z. Zhang\fnref{fn1}}

\author[2]{Adrian J. Bevan\corref{cor1}%
  }
\ead{a.j.bevan@qmul.ac.uk}

\author[1]{M. Bullough}

\author[3]{J. Dopke}

\author[2]{J. Mistry}

\author[2]{S. Zenz}

 \cortext[cor1]{Corresponding author}

\affiliation[1]{organization={Micron Semiconductor Ltd.}, 
                 addressline={83 Marlborough Rd},
                 postcode={BN15 8SJ}, 
                 city={Lancing}, 
                 country={UK}}

\affiliation[2]{organization={Queen Mary University of London}, 
                 addressline={School of Physical and Chemical Sciences, Mile End Road},
                 postcode={E1 4NS}, 
                 city={London}, 
                 country={UK}}
\affiliation[3]{organization={STFC, Rutherford Appleton Laboratory}, 
                 addressline={Harwell},
                 postcode={OX11 0QX}, 
                 city={Didcot}, 
                 country={UK}}

\fntext[fn1]{Now at the University of Sussex.}

\begin{abstract}
For many years there has been an aspiration within the community to develop curved silicon detectors for particle physics applications. We present the results from $10\times 10$cm low mass support modules as a part of the “ZeroMass” project that aims to minimise the material budget for tracking and vertexing systems for future colliders. We use 50 $\mu$m thick DC coupled strip sensors from Micron Semiconductor Ltd., with a carbon composite support frame. Our current module demonstrators use a radius of curvature of 15cm, typical of that used for the outer parts of large pixel systems, or the inner part of strip trackers and the outer part of large radii vertex detectors. The material budget obtained varies from an $X_0$ of 0.05\% in the active area to 0.62\% in the support structure, with an average of 0.28\%. There is further scope for material budget reduction in applying the concept and methods to large instruments for future detector systems, which we also discuss.
\end{abstract}


\end{frontmatter}


\section{Introduction}
\label{sec:intro}
Ultra thin silicon below about 75$\mu$m is flexible. An advantage of this is that if we conform the silicon to a particular shape, then the crystal becomes part of a rigid mechanical structure.  For passively cooled detectors there is the opportunity to explore detector concepts of cylindrically curved large area sensors bent into shape and kept rigid using a light weight mechanical structure (e.g. carbon fibre-reinforced plastic, CFRP) to build a lightweight large area tracker.  For this concept to be viable we first have to validate that we are able to make and operate individual sensor. To this end we have  partnered with Micron Semiconductor Ltd. in Lancing, UK to use a bespoke variant of their TTT10 32 channel DC coupled strip detector~\footnote{\url{https://www.micronsemiconductor.co.uk/product/ttt10/}}.

If the radius of curvature of ultrathin silicon [a flexible thin film (TF)] is large enough to satisfy the Griffith criterion~\cite{Griffith1921}, then the sensor can be formed into a cylindrical (or spherical) shape without cracking.  We calculate a minimum physical radius of 4mm for our sensors.  Successful assembly of a curved module relies on other constraints, including tooling to assemble a module with such a small curvature.  We have developed both manual and vacuum assembly approaches; the former has the advantage of being very reproducible, but the disadvantage of being reliant on the individual performing assembly. Vacuum tooling, by design, increases the placement precision, but introduces other challenges, including complexity, to the assembly process.

\section{Mechanical tokens}
\label{sec:modules}

As a part of the Arachnid collaboration, that focused on the testing of CMOS MAPS for collider experiments~\cite{Arachnid:2013hee}, one of the authors constructed mechanical models of curved detector modules using 50$\times$20mm silicon dies~\cite{Wilson:2013}.  These objects were used to gain familiarity with handling 50$\mu$m silicon, and  have demonstrated the ability to curve 50\mum thick silicon down to a radius of 13mm.  A radius of 25mm was used to make mechanical structures with curved silicon to study stability over time.  TF theory predicts left undisturbed these should remain intact as defects can not propagate.  We observe that curved silicon tokens made in 2012 remain intact today - more than a decade later, as expected. Thus we can construct curved silicon detectors that last at least a decade.

\section{Detector modules}
\label{sec:modules}

\begin{figure}[!ht]
\centering
\includegraphics[width=9.0cm]{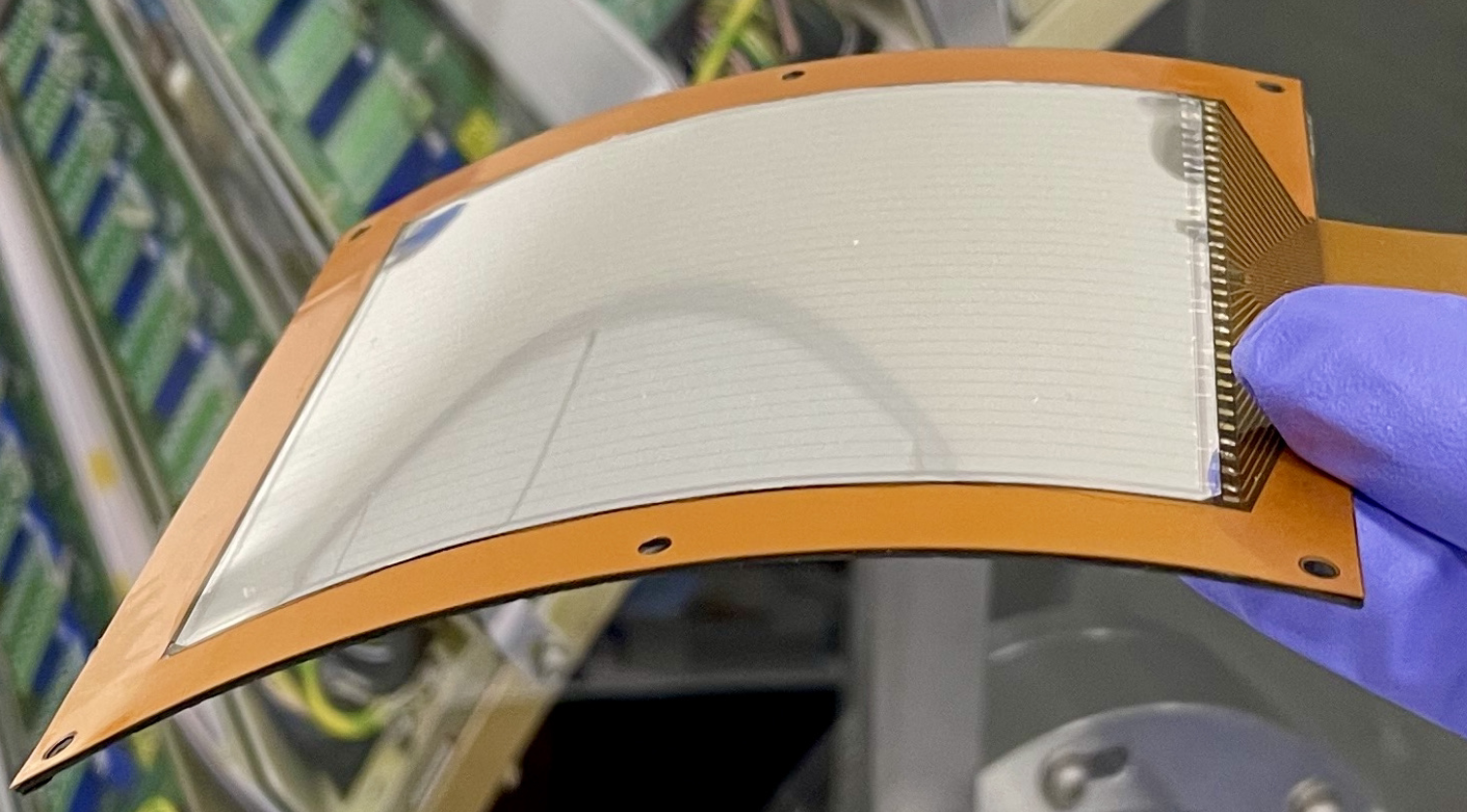}
\caption{A curved silicon module constructed from a TTT10 sensor.  The CFRP support frame overlaps with the edge of the chip by 2mm.}\label{module}
\end{figure}

Over the past few years our team has assembled flat and curved silicon modules, where the focus has been on using a 150mm radius of curvature.  Figure~\ref{module} shows one of our modules with a CFRP support frame. 
The active area of the TTT10 sensor is 95.97$\times$95.97 mm.

\vspace{-0.5cm}

\section{Piezoresistance}
\label{sec:piezo}

One of the aims of this project was to understand how sensor performance changed when stress was introduced.  We observe an increase in dark current when we curve the silicon to form modules. This increase is typically a few nA, and does not affect our ability to use the sensors to measure $\alpha$ radiation from a $\,^{241}$Am source.  The change in dark current (factor of 2 increase) between a flat and curved sensor is not accompanied by a change in capacitance, and as a result we conclude that this is a piezoresistance effect.  See for example~\cite{piezo}.

\begin{figure}[!ht]
\centering
\includegraphics[width=9.0cm]{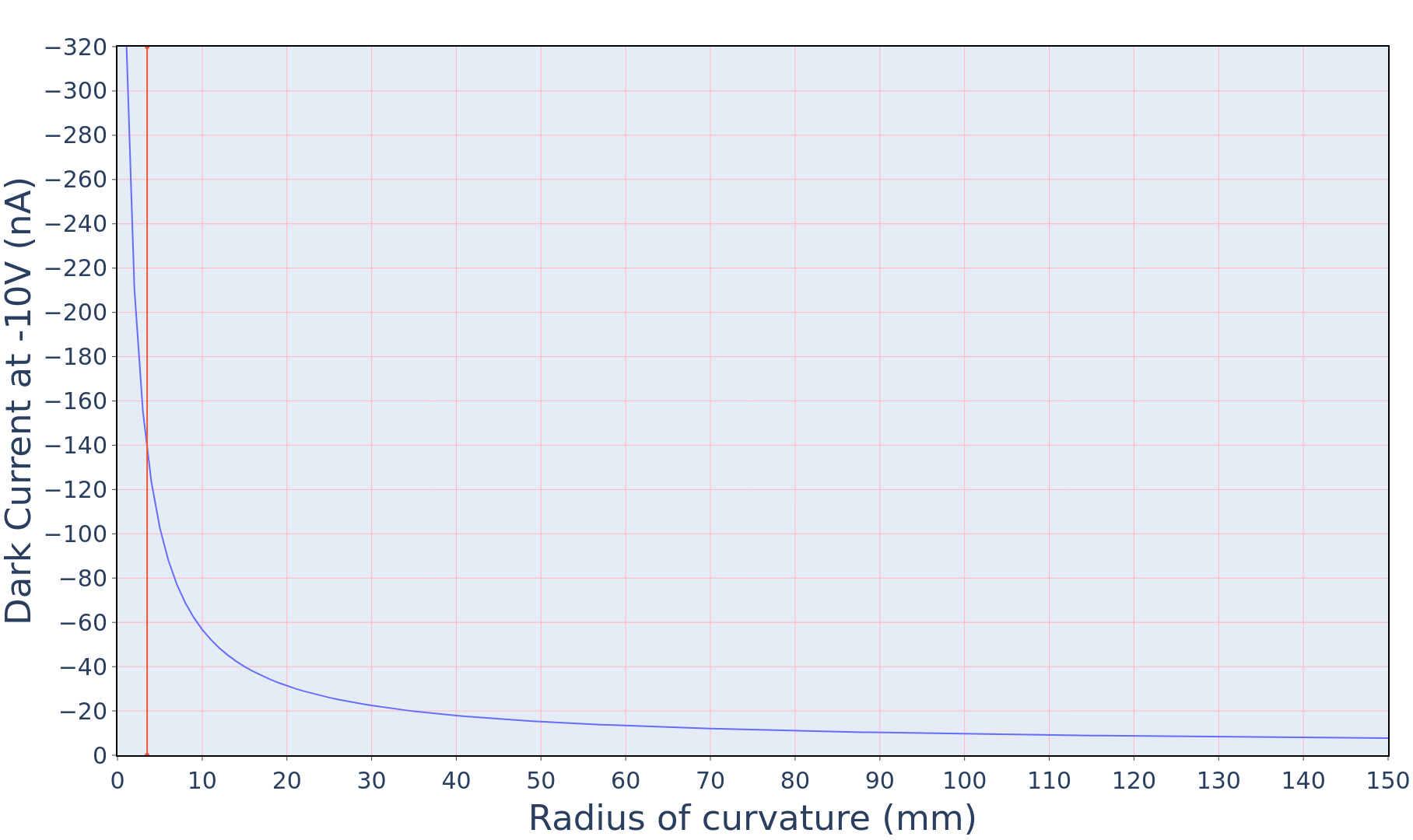}
\caption{The predicted dark current as a function of radius of curvature for one of our TTT10 sensors used in a curved module. The vertical line indicates the physical limit of 4mm bending radius calculated for the Griffith criterion.}\label{fig:piezo}
\end{figure}

We use strain guages to calibrate the characteristic response curve for this effect. This enables us to determine the expected change in dark current as a function of the radius of curvature, shown in Figure~\ref{fig:piezo}.  This indicates dark currents of a few tens of nA would be generated for radii down to a few cm. 

\section{Sensor response}
\label{sec:response}

\begin{figure}[!ht]
\centering
\includegraphics[width=9.0cm]{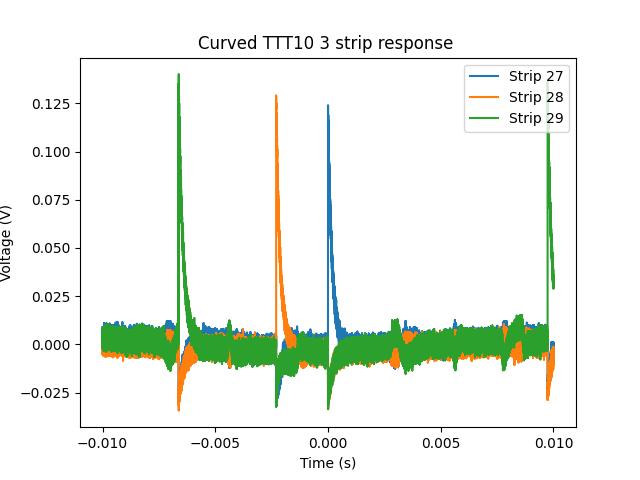}
\caption{The response of 3 adjacent strips in the curved sensor module to $\alpha$ radiation.}\label{signals}
\end{figure}

The response of strips in a curved sensor to $\alpha$ radiation from a $\,^{241}$Am radiation is shown in Fig.~\ref{signals}.  The figure shows the response of 3 adjacent strips that are read out with Cremat CR110 amplifiers\footnote{https://www.cremat.com/home/charge-sensitive-preamplifiers/} feeding into a Tektronix MSO44 oscilloscope.  The shaping time of the amplifier is 140$\mu$s, giving the characteristic fall off observed.  The bias voltage used to collect this data is 5V, and the amplifiers were powered using a 9V supply.

\section{Summary and future work}
\label{sec:summary}

This work demonstrates that it is possible to make large area cylindrically curved modules.   The motivation of this work using strip sensors was to study the effect of bending on the signal response.  We have demonstrated it is possible to understand piezoresistance induced changes. We have identified industrial challenges that may benefit from this work, specifically radioactive waste characterisation.  We are now working with industry partners on this problem, which will enable us to explore the validity of the extrapolation in Fig.~\ref{fig:piezo}.
We are reverting to CMOS and have been working with 50$\mu$m thick ATLASpix3.1 chips in order to develop the work using CMOS.


  \bibliographystyle{elsarticle-num-names} 
  \bibliography{bibfileBevanPisaMeetingProcsZMD}

\end{document}